\documentclass[sigconf]{acmart}
\usepackage{multirow}
\usepackage{svg}
\usepackage{subcaption}
\usepackage{threeparttable}
\usepackage[normalem]{ulem}
\usepackage{enumitem}
\usepackage{float}
\useunder{\uline}{\ul}{}
\newtheorem{mydef}{Definition}
\AtBeginDocument{%
  \providecommand\BibTeX{{%
    \normalfont B\kern-0.5em{\scshape i\kern-0.25em b}\kern-0.8em\TeX}}}
\setcopyright{acmlicensed}
\copyrightyear{2018}
\acmYear{2018}
\acmDOI{XXXXXXX.XXXXXXX}

\acmConference[Conference acronym 'XX]{Make sure to enter the correct
  conference title from your rights confirmation email}{June 03--05,
  2018}{Woodstock, NY}
\acmISBN{978-1-4503-XXXX-X/18/06}

\begin{document}

%%
%% The "title" command has an optional parameter,
%% allowing the author to define a "short title" to be used in page headers.
% \title{Cross-view Graph Self-supervised Learning for Group Identification via Transitional Hypergraph Convolution}
% \title{Unified Multitask pretraining for Recommendation via Hypergraph with Transitional Attention}
\title{Instruction-based Hypergraph Pretraining}

%%
%% The "author" command and its associated commands are used to define
%% the authors and their affiliations.
%% Of note is the shared affiliation of the first two authors, and the
%% "authornote" and "authornotemark" commands
%% used to denote shared contribution to the research.
\author{Mingdai~Yang}
\affiliation{%
  \institution{University of Illinois at Chicago}
  \city{Chicago}
  \country{USA}}
\email{myang72@uic.edu}

\author{Zhiwei~Liu}
\affiliation{%
  \institution{Salesforce AI Research}
  \city{Palo Alto}
  \country{USA}
}
\email{zhiweiliu@salesforce.com}

% \author{Liangwei~Yang, Xiaolong~Liu, Chen~Wang}
% \affiliation{%
%   \institution{University of Illinois at Chicago}
%   \city{Chicago}
%   \country{USA}}
% \email{{lyang84, xliu262, cwang266}@uic.edu}

\author{Liangwei~Yang}
\email{lyang84@uic.edu}
% \author{Xiaolong~Liu}
% \email{xliu262@uic.edu}
\affiliation{%
  \institution{University of Illinois at Chicago}
  \city{Chicago}
  \country{USA}}

\author{Xiaolong~Liu}
\email{xliu262@uic.edu}
\author{Chen~Wang}
\email{cwang266@uic.edu}
\affiliation{%
  \institution{University of Illinois at Chicago}
  \city{Chicago}
  \country{USA}}

\author{Hao Peng}
\affiliation{%
   \institution{School of Cyber Science and Technology, Beihang University,}
   \country{Beijing, China}\\
   \institution{Yunnan Key Laboratory of Artificial Intelligence, Kunming University of Science and Technology,}
   \country{Kunming, China}}
\email{penghao@buaa.edu.cn}
\authornote{Corresponding author}

\author{Philip S.~Yu}
\affiliation{%
  \institution{University of Illinois at Chicago}
  \city{Chicago}
  \country{USA}}
\email{psyu@uic.edu}

%%
%% By default, the full list of authors will be used in the page
%% headers. Often, this list is too long, and will overlap
%% other information printed in the page headers. This command allows
%% the author to define a more concise list
%% of authors' names for this purpose.

%%
%% The abstract is a short summary of the work to be presented in the
%% article.
\begin{abstract}
Pretraining has been widely explored to augment the adaptability of graph learning models to transfer knowledge from large datasets to a downstream task, such as link prediction or classification. However, the gap between training objectives and the discrepancy between data distributions in pretraining and downstream tasks hinders the transfer of the pretrained knowledge. Inspired by instruction-based prompts widely used in pretrained language models, we introduce instructions into graph pretraining. In this paper, we propose a novel pretraining framework named \textbf{I}nstruction-based \textbf{H}ypergraph \textbf{P}retraining. 
To overcome the discrepancy between pretraining and downstream tasks, text-based instructions are applied to provide explicit guidance on specific tasks for representation learning. Compared to learnable prompts, whose effectiveness depends on the quality and the diversity of training data, text-based instructions intrinsically encapsulate task information and support the model to generalize beyond the structure seen during pretraining.
To capture high-order relations with task information in a context-aware
manner, a novel prompting hypergraph convolution layer is devised to integrate instructions into information propagation in hypergraphs. Extensive experiments conducted on three public datasets verify the superiority of IHP in various scenarios.
\end{abstract}

%%
%% The code below is generated by the tool at http://dl.acm.org/ccs.cfm.
%% Please copy and paste the code instead of the example below.
%%
\begin{CCSXML}
<ccs2012>
   <concept>
       <concept_id>10002951.10003317.10003338</concept_id>
       <concept_desc>Information systems~Retrieval models and ranking</concept_desc>
       <concept_significance>500</concept_significance>
       </concept>
 </ccs2012>
\end{CCSXML}

\ccsdesc[500]{Information systems~Retrieval models and ranking}
%%
%% Keywords. The author(s) should pick words that accurately describe
%% the work being presented. Separate the keywords with commas.
\keywords{Graph Pretraining; Hypergraph Learning}

%% A "teaser" image appears between the author and affiliation
%% information and the body of the document, and typically spans the
%% page.

%%
%% This command processes the author and affiliation and title
%% information and builds the first part of the formatted document.
\maketitle

\section{Introduction}
%\textbf{Introduce graphs and GNNs, Introduce Graph pretraining: Knowledge transfer. Challenge: Gap between pretext and downstream tasks; suffer from forgetting with small-scale downstream}

The burgeoning field of graph-based deep learning has witnessed remarkable advancements, with applications spanning diverse domains such as social networks, bioinformatics, and recommender systems. Graph Neural Networks (GNNs) have emerged as powerful tools for capturing intricate relationships and dependencies within graph-structured datasets. To transfer knowledge from large and diverse datasets to a downstream task with limited data, pretraining has been widely explored to augment the adaptability of GNNs~\cite{gptgnn,gcc}. 
% Similar to other neural network architectures benefitting from pretraining, graph models exposed to a variety of structures and features during pretraining are enabled to develop a more robust understanding of graph data. 
However, pretraining also presents several critical issues.
Self-supervised learning for pretraining graph models~\cite{gcc,attrimask} addresses the lack of labeled data in the pretext stage, but the training objective gap between the constructed pretext and dedicated downstream tasks hinders the efficient transfer of pretrained knowledge~\cite{gppt}. 
Moreover, when labeled data is available during the pretraining stage, the data distribution discrepancy between pretraining and downstream datasets degrades the performance~\cite{afec}. In addition, pretrained can suffer from catastrophic forgetting~\cite{forgetting}, resulting in poor generalization ability, especially within small-scale downstream datasets.

%\textbf{Introduce Graph Prompting; why better than the previous pretrain-finetune pattern? Limitation of current works: (a) randomly initialized learnable prompt tokens not related to semantic instructions/specific tasks, and (b) exact same graph for pretext and downstream tasks. Unable to deal with a downstream graph extended with unseen nodes/edges}
Prompt learning, which has been prevalently engaged in Pretrained Language Models (PLMs)~\cite{promptsurveynlp, gpt3}, has shown remarkable success in dealing with the aforementioned issues of pretraining. Nonetheless, it is far from straightforward to apply prompts in graphs.  In contrast with PLMs, the format of instructions as plain text does not naturally align with the graph-structure data.
On the one hand, from a static perspective, graph prompts are supposed to be applied for a certain part of the graph to guide the model depending on specific graph queries. Some
specific tasks, such as group identification, can correspond to a
relationship among multiple entities. However, traditional graphs
are limited in their ability to represent complex relationships, as
each edge can only connect two nodes. On the other hand, from a
dynamic perspective, prompts need to participate in the context-
aware information propagation of the graph model to help capture
contextual dependencies among relationships and entities.
Nevertheless, dyadic edges in traditional graphs can only propagate
information through simple pairwise relationships, which is neither
flexible nor efficient. Therefore, we believe that the hypergraph, where each hyperedge is able to simultaneously connect multiple nodes~\cite{hgnn,ZhouHS06}, is a better structure for prompt-based pretraining in graph-based tasks.

%The aim of prompt learning is to utilize the model's pretrained capabilities to perform specific downstream tasks as guided by the prompt. This idea has also been introduced to graph-based tasks. For a pretrained graph model, prompts can reformulate downstream tasks in line with the pretraining task to overcome the discrepancy between them. 
% \textbf{---Under Revision---}

However, it is challenging to design the prompt for hypergraph pretraining. Most pioneering works in prompt-based graph pretraining design prompt functions with learnable prompt vectors, which are initialized randomly~\cite{allinone,graphprompt} or based on pretrained node representations~\cite{gppt}. Such prompts are not related to semantic instructions for specific tasks, and the effectiveness heavily depends on the quality and diversity of the finetuning data in downstream tasks. The learnable prompt vectors can bring uncertainty in outcomes and the risk of poor generalization, particularly when finetuning data is biased or scarce. In addition,
most previous works freeze all parameters except learnable prompts during downstream tasks~\cite{allinone,graphprompt,gppt}. However, this paradigm is suboptimal if abundant unseen nodes appear in the downstream task. The frozen nodes pretrained by pretext tasks cannot offer a direct reference to represent unseen nodes in the downstream task, and the uncertainty from learnable prompts can be exacerbated with those unseen data.
These limitations hinder the implementation of these pretraining frameworks in graph-based applications, given that new nodes frequently appear in dynamically evolving graph structures such as social networks, biological networks, and recommender systems.

%\textbf{Why apply instruction-based prompt to graph: (a)provide reliable ground-truth reference for pretext and downstream tasks; (b)unseen nodes/edges in downstream tasks: easily solved by instruction-based prompt}
In light of the above limitations in the existing prompt learning methods for graph-based tasks, it is necessary to design prompts that can provide explicit guidance to the model, helping it focus on specific aspects of the task. To this end, instruction-based prompts represent a more advantageous solution~\cite{instructor,gpt3,graphgpt}. Such prompts guide the model through explicit instructions, which are able to accurately describe specific task requirements related to graph data. By training the model with a variety of instructions, it can adapt to different graph-related queries, improving its generalization across a range of scenarios. In the situation with biased or scarce data, compared to learnable prompts which may overfit the limited examples, instructions allow for the incorporation of domain-specific knowledge that might not be present in the data itself.  Instruction-based prompts are also beneficial for addressing unseen nodes in downstream tasks. Rather than relying solely on previously seen examples, text-based instructions provide side information on the new nodes appearing in the graph. Meanwhile, this adaptation to the unseen graph structure requires no extensive computational resources for re-training prompts.

%\textbf{How to apply instruction-based prompt to graph: Hyperedges; flexibility of hyperedges - suitable for various graph tasks; method - prompt task-related information as an instruction to task-related nodes during information propagation}

%Despite the numerous benefits of instruction-based prompts, it is challenging how to apply them to graph tasks. In contrast with PLMs, the format of instructions as plain text does not naturally align with the graph-structure data. On the one hand, from a static perspective, instructions are supposed to be prompted for a certain part of the graph, which may include multiple nodes, to guide the model depending on specific graph queries. Some specific tasks, such as group identification, can correspond to a relationship among multiple entities. However, traditional graphs are limited in their ability to represent complex relationships, as each edge can only connect two nodes. On the other hand, from a dynamic perspective, instructions need to participate in the context-aware information propagation of the graph model to help capture contextual dependencies among relationships and entities. Under ideal circumstances, instructions improve the flow of information through multiple related nodes during the information propagation. However, dyadic edges in traditional graphs can only propagate information through simple pairwise relationships, which is neither flexible nor efficient. 

\begin{figure}[t!]
\centering
    \includegraphics[width=0.95\linewidth]{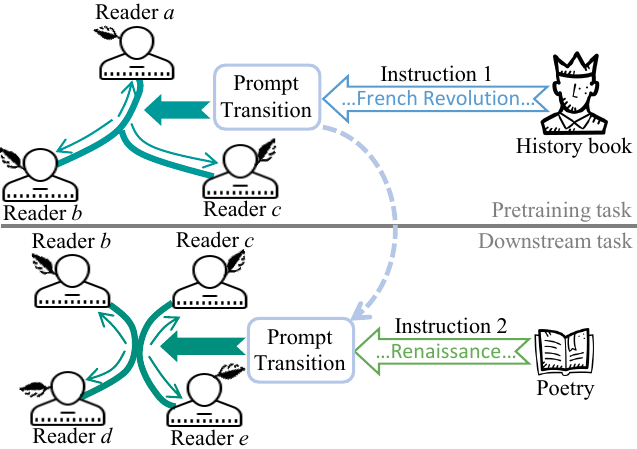}
  \caption{A toy example of how related instructions benefit pretext and downstream tasks with high-order relations.}
  \label{fig:overview}
\end{figure}
% Therefore, we believe that the hypergraph is a better structure for pretraining with instructions in graph-based tasks. 
% In hypergraphs, each hyperedge is able to connect multiple nodes simultaneously instead of only two nodes~\cite{hgnn,ZhouHS06}.
% We show an example of instruction-based pretraining with high-order relations in Fig.~\ref{fig:overview}. If three history enthusiasts read a history book about the French Revolution, we can prompt the model with this instruction through pretraining. In the following downstream task of promoting poetry from the Renaissance period to readers, the model can accordingly locate the readers similar to these three history enthusiasts as the target audiences according to the instructions. 
% In both pretraining and downstream tasks, the high-order relationships can be inherently captured by hyperedges.
% So far, how to integrate the instruction into hypergraph learning to capture such high-order relationships is still under-researched.

% Within instruction-based pretraining, the other challenge is how to deal with the unseen data in downstream graph tasks. In contrast to randomly initialized learnable prompts, text-based instructions explicitly convey task-related information. 
%However, how to effectively leverage such information from instructions in graph pretraining remains insufficiently explored.

In this paper, we propose a novel framework called Instruction-based Hypergraph Pretraining (IHP). In this framework, high-order relations are represented as hyperedges, and the dependencies among nodes are captured under the guidance of instructions.
We show an example of how to integrate instructions into hypergraph pretraining in Fig.~\ref{fig:overview}. If three history enthusiasts read a history book about the French Revolution, we can use one hyperedge to connect them and prompt the hyperedge with instruction through pretraining. In the following downstream task of promoting poetry from the Renaissance period to readers, the model can accordingly locate the readers similar to these three history enthusiasts as the target audiences according to the instruction. 
To learn node representations under the guidance of instructions, we design a Prompt Hypergraph Convolution (PHC) layer. This PHC layer endows the framework with the ability to be prompted by text-based instructions through hyperedges. Within the hypergraph learning framework, each task-related instruction transformed into prompts is precisely directed to the respective nodes interconnected by the hyperedge. Participating in hypergraph learning, instructions enable the framework to capture high-order relations with task information in a context-aware manner. Furthermore, we design an instruction-based finetuning paradigm. In the downstream task, we freeze the prompt transformation process and update the representations of seen and unseen nodes according to the instructions, ensuring that relationships among nodes are captured in the context of the overall graph. Node representations are updated with different adaptation intensities to achieve a balance between retraining prior knowledge and adapting efficiently for downstream tasks.

The main contributions of this paper are summarized as follows:
\begin{itemize}[leftmargin=*]
    \item We design a novel framework IHP, an instruction-based pretraining framework based on hypergraphs. To the best of our knowledge, this is the first pretraining framework to leverage instructions to capture high-order relations for graph tasks.
    \item We proposed a novel PHC layer to prompt text-based instructions into hyperedges. This PHC layer allows instruction information to participate in information propagation during hypergraph convolution, enhancing the flexibility of the hypergraph learning and the generalization of the pretrained model.
    \item We conduct extensive experiments on three real-world datasets. The marked enhancement observed in the performance of IHP in various scenarios underscores its preeminence as an instruction-based pretraining framework.

\end{itemize}

\section{Related works}
\subsection{Hypergraph Learning}
Interactions in most real-work networks are more complex than dyadic edges used in traditional graphs, in which cases a hypergraph provides a more expressive structure to represent such relations~\cite{LungGS18,JiaZDP21,FeigeFIILS15}. 
% For instance, one paper can have multiple authors with varying degrees of contribution in co-authorship networks~\cite{LungGS18}, the interest of a group discriminatively depends on its members in online communities~\cite{JiaZDP21}, and items have complements and substitutes reflected by co-purchasement in e-commercial platforms~\cite{FeigeFIILS15}.Due to the ability of hypergraph to capture complex high-order dependencies through hyperedge-node connections, the application of hypergraph learning has proven successful in various graph-based tasks.
Beyond the basic hypergraph convolution and attention neural networks\cite{hgnn,BaiZT21}, hypergraph learning demonstrates adaptability for customization across a variety of graph-based tasks. DHCF~\cite{dhcf} employs residual connections to effectively capture hybrid multi-order correlations in the user-item graph, simultaneously considering aggregated representations from the original graph and the hypergraph. MHCN~\cite{mhcn} introduces a multi-channel hypergraph convolutional network that leverages different types of high-order relations among users to predict social links. Seq-HyGAN~\cite{SaifuddinMTIA23} proposes sequential hypergraph attention to capture dependencies between extracted subsequences for sequence classification.
To the best of our knowledge, no previous method has applied hypergraph learning for instruction-based pretraining.

\subsection{Graph Pretraining and Prompting}
Graph learning has attracted interest owing to the prevalence of graph-structured data in various fields. To improve the learning efficacy of such graphs, researchers have been investigating graph pretraining utilizing unlabeled graph data~\cite{Jiang21,l2pgnn,attrimask}.
% PT-HGNN~\cite{Jiang21} introduces node-level and schema-level pretraining to preserve heterogeneous semantic and structural properties through contrastive learning.
% L2P-GNN~\cite{l2pgnn} proposes a self-supervised GNN pretraining strategy that adaptively addresses the divergence between pretraining and finetuning objectives. 
GCC~\cite{gcc} applies contrastive learning to capture the universal topological properties across multiple networks.
GPT-GNN~\cite{gptgnn} introduces a graph generation task during pretraining to capture the characteristics of the graph through the utilization of node features.

With the prevalence of prompt learning, prompting frameworks are applied to mitigate the training objective gap and catastrophic forgetting. A pioneering work GPPT~\cite{gppt} designs a pairwise prompting function generating learnable token pairs from standalone nodes for downstream tasks. GraphPrompt~\cite{graphprompt} multiplies each node embedding by a prompt vector into graph aggregation functions for different tasks. Sun et al.~\cite{allinone} introduce a prompting framework that reformulates nodes and edges to induced subgraphs to narrow the gap between different tasks during pretraining and finetuning. However, these works are limited to homogeneous graphs, and learnable prompts provide no explicit semantic clues on graph structures and tasks. On the contrary, our method leverages instruction-based prompts based on plain text instead of learnable prompts. With the guidance of instructions delivered by hyperedges, IHP is able to capture the high-order relations among different nodes.

\subsection{Graph Learning with Text Information}
Instructions can be regarded as additional text information to the graph structure. Incorporating text into graph learning offers extra contextual understanding and semantic insights, especially when the downstream graph structure is sparse. With the breakthroughs in PLMs, it is straightforward to leverage PLMs and GNNs for textual graph representation following a cascaded architecture~\cite{graphsage}. In the cascaded architecture, text information is first encoded by text encoders and then aggregated by graph models to obtain the final representation~\cite{Jason21,Chaozhuo21,Di21}. In contrast with cascaded architectures, GraphFormers~\cite{graphformer} and EdgeFormers~\cite{edgeformer} propose GNN-nested Transformers to jointly encode text and node features in textual graphs. 
% Besides homogeneous graphs with text~\cite{Jason21,graphformer,edgeformer}, some previous studies focus on leveraging text information in heterogeneous graphs for specific text-related tasks, 
% such as taxonomy construction~\cite{Jingbo20} and hypernymy discovery~\cite{Yu19}. For the generic representation learning task, SHNE~\cite{shne} and Heterformer~\cite{heterformer} capture structure features with textual signals by the cascade architecture and the GNN-nested Transformer, respectively.
Different from previous works straightforwardly incorporating text encoding in heterogeneous graph learning~\cite{shne, heterformer}, IHP focuses on leveraging instructions to transfer learning capabilities from pretext tasks to downstream tasks. By pretraining with instruction-based prompts, IHP enables the model to develop a generalized understanding of graph properties, which is beneficial when dealing with unseen graph data.
\section{Preliminaries}
\subsection{Problem Formulation}
Let the graph for the pretraining task be $G=(V, E)$ where $V$ and $E$ are sets of nodes and edges, respectively. Similarly, let the graph for the downstream task be $G'=(V', E')$, and the two instruction sets be $I_p$ and $I_d$ for the pretraining task and the downstream task. Our target is to pretrain the representations of the overlapping nodes, which are defined as target nodes $\mathcal{V}_\text{t}=V\cap V'$. For example, the target nodes can be the overlapping molecules of different proteins in protein classification, or the overlapping users purchasing items of different domains in cross-domain recommendation.  Given the target node set $\mathcal{V}_\text{t}$, we define the nodes other than target nodes in the pretraining task as pretraining context nodes $\mathcal{V}_\text{c}=V\setminus V_t$ in the pretraining task, and the unseen nodes in the downstream task as downstream context nodes $\mathcal{V}_\text{c}'=V'\setminus V_t$. Besides, we define the descriptions of pretraining nodes as $\mathcal{A}$ and the descriptions of downstream nodes as $\mathcal{A}'$. The descriptions of different tasks are denoted as $\mathcal{T}$, with each task related to arbitrary nodes in graphs. These descriptions are used to construct instructions if available.
\subsection{Hypergraph}
%We introduce two hypergraphs, target hypergraph and context hypergraph, for the representation learning of target nodes and context nodes during pretraining.
\begin{mydef}
\textbf{(Hypergraph)}. 
A hypergraph is defined as $\mathcal{H} = (\mathcal{V},\mathcal{E})$, where $\mathcal{V}$ denotes the node set and $\mathcal{E}$ represents the edge set.  An incidence matrix $\mathbf{H}\in\{0,1\}^{|\mathcal{V}|\times|\mathcal{E}|}$ is used to represent connections among nodes and hyperedges in the hypergraph. 
\end{mydef}
Compared with an edge connecting only two nodes in a traditional graph, a hyperedge connects multiple nodes simultaneously in a hypergraph. Hence, hypergraph naturally possesses the ability to model higher-order connections.

% \begin{mydef}
% \textbf{(Target Hypergraph)}. 
% Given the target node set $\mathcal{V}_\mathbf{t}$, the pretraining target hypergraph is denoted as $\mathcal{H}_\mathbf{t} = (\mathcal{V}_\mathbf{t},\mathcal{E}_\mathbf{t})$, where $\mathcal{E}_\mathbf{t}$ are hyperedges connecting target nodes in the pretraining task. An incidence matrix $\mathbf{H}_\mathbf{t}\in\{0,1\}^{|\mathcal{V}_\mathbf{t}|\times|\mathcal{E}_\mathbf{t}|}$ is used to represent connections among the target nodes in the pretraining task. 
% \end{mydef}
% \begin{mydef}
% \textbf{(Context Hypergraph)}. 
% Given the pretraining context node set $\mathcal{V}_\mathbf{c}$, the pretraining context hypergraph is denoted as $\mathcal{H}_\mathbf{c} = (\mathcal{V}_\mathbf{c},\mathcal{E}_\mathbf{c})$, where $\mathcal{E}_\mathbf{c}$ are hyperedges connecting context nodes in the pretraining task. An incidence matrix $\mathbf{H}_\mathbf{c}\in\{0,1\}^{|\mathcal{V}_\mathbf{c}|\times|\mathcal{E}_\mathbf{c}|}$ is used to represent connections among the context nodes in the pretraining task.
% \end{mydef}
% Similarly, a target hypergraph $\mathcal{H}_\text{c}' = (\mathcal{V}_\text{t},\mathcal{E}_\text{t}')$ and a context hypergraph $\mathcal{H}_\text{c}' = (\mathcal{V}_\text{c}',\mathcal{E}_\text{c}')$ are defined for the downstream task with their corresponding incidence matrices $\mathbf{H}_\text{t}'$ and $\mathbf{H}_\text{c}'$.

\section{Method}
\begin{figure*}[h]
  \centering
    \includegraphics[width=\linewidth]{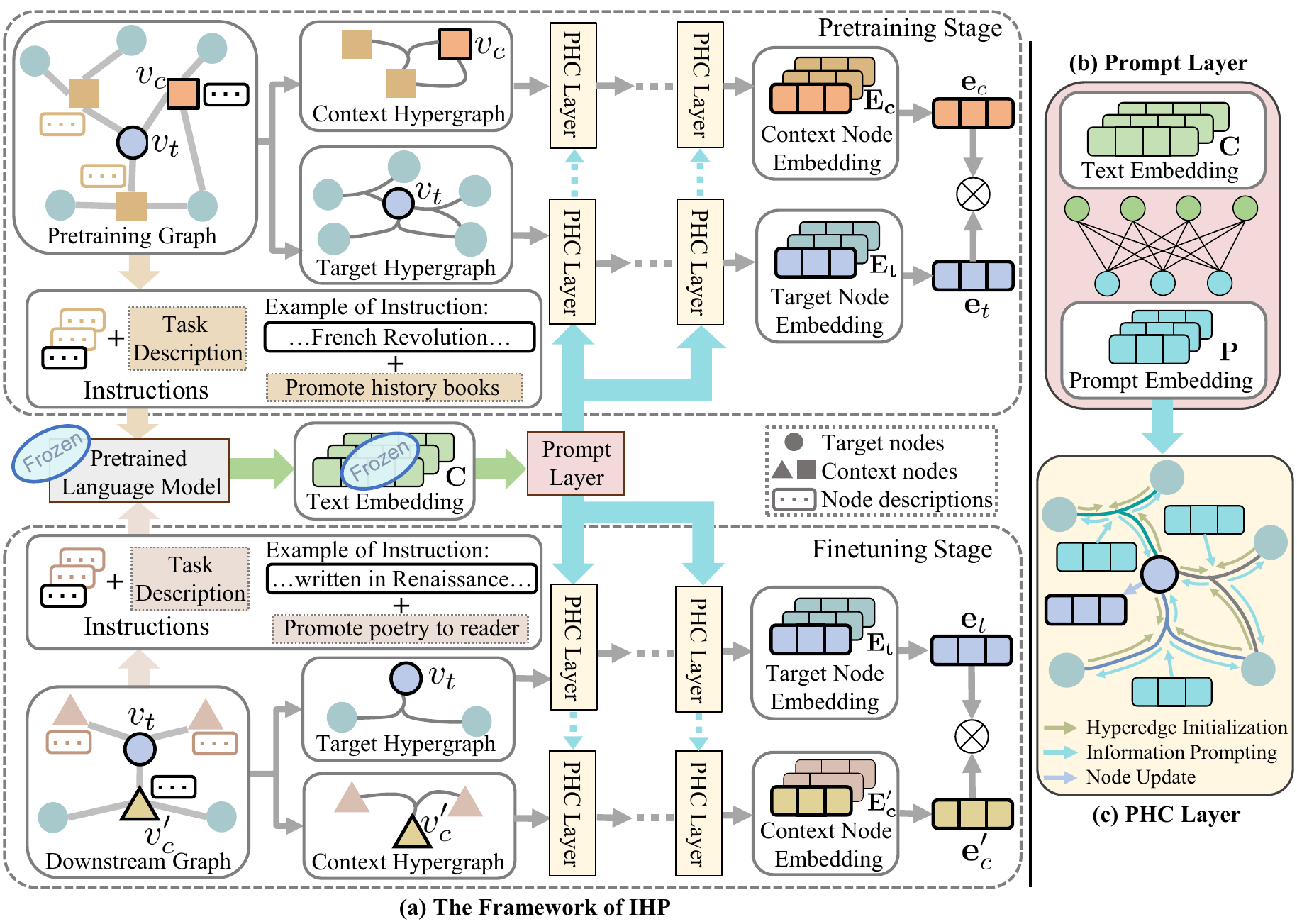}
  \caption{(a) The overall framework of IHP. The target nodes are the nodes existing in both pretraining and finetuning stages, and other nodes are defined as context nodes; (b) The illustration of the Prompt Layer; (c) The illustration of the PHC Layer.}
%   The details of Tripartite Convolution, Factorized Attention and Ranking Prediction are discussed in Section~\ref{TriConv}, Section~\ref{FacAtt} and Section~\ref{RanPre} respectively.
  \label{fig:IHP}
\end{figure*}

In this section, we present the proposed IHP framework with The illustration demonstrated in
Figure~\ref{fig:IHP}. We start by introducing all the embeddings to be pretrained in this framework. Thereafter, we define two types of hypergraphs and explain how to construct corresponding hyperedges in these hypergraphs. Specifically, we adopt a PHC layer to prompt instructions into hyperedges during hypergraph learning in both pretraining and finetuning stages. We apply an instruction-based finetuning paradigm to update both seen and unseen
nodes in the downstream task.

\subsection{Embedding Layer}
We maintain an embedding layer $\mathbf{E}\in \mathbb{R}^{(|{\mathcal{V}_\text{t}}|+|\mathcal{V}_\text{c}|)\times d}$ during pretraining, where $d$ is the feature dimension and columns represent all trainable node embeddings. The initial target node embeddings are denoted as $\mathbf{E}_\text{t}$ and the initial context node embeddings are denoted as $\mathbf{E}_\text{c}$. 
Since target nodes exist in pretraining and downstream tasks, prior knowledge learned from the pretraining tasks can be preserved in $\mathbf{E}_\text{t}$ and then leveraged in the downstream tasks.

\subsection{Hypergraph Construction}\label{sec:hyperedge}
% 3.2 section
We construct the target hypergraph and the context hypergraph based on the original graph.
Compared to a graph where edges only connect two nodes, the advantage of a hypergraph is that hyperedges can simultaneously connect multiple nodes as the objectives of an instruction. 
According to the type of nodes in the graph, we construct two hypergraphs based on four basic types of hyperedges as below:
% \begin{mydef}
% \textbf{(Target Hypergraph)}. 
% Given the target node set $\mathcal{V}_\mathbf{t}$, the pretraining target hypergraph is denoted as $\mathcal{H}_\mathbf{t} = (\mathcal{V}_\mathbf{t},\mathcal{E}_\mathbf{t})$, where $\mathcal{E}_\mathbf{t}$ are hyperedges connecting target nodes in the pretraining task. An incidence matrix $\mathbf{H}_\mathbf{t}\in\{0,1\}^{|\mathcal{V}_\mathbf{t}|\times|\mathcal{E}_\mathbf{t}|}$ is used to represent connections among the target nodes in the pretraining task. 
% \end{mydef}

\begin{itemize}[leftmargin=*]
    \item \textbf{Target Hypergraph}. The target hypergraph $\mathcal{H}_\mathbf{t}$ consists of target nodes $\mathcal{V}_\mathbf{t}$ and hyperedges $\mathcal{E}_\mathbf{t}$. Each target-target hyperedge is used to connect a target node with its one-hop target node neighbors. For each group of target nodes connected to one context node, a target-context hyperedge is used to connect them in the hypergraph. Its incidence matrix is denoted as $\mathbf{H}_\text{t}$.
    \item \textbf{Context Hypergraph}. Similar to the target hypergraph, the context hypergraph $\mathcal{H}_\mathbf{c}$ consists of context nodes $\mathcal{V}_\mathbf{c}$ and hyperedges $\mathcal{E}_\mathbf{c}$. Each context-context hyperedge is used to connect a context node with its one-hop context node neighbors. For each group of context nodes connected to one target node, a context-target hyperedge is used to connect them simultaneously in the hypergraph. Its incidence matrix is denoted as $\mathbf{H}_\text{c}$.
\end{itemize}
In other words, we use the two hypergraphs to ensure the distinction between the target nodes and the context nodes.
In this way, the framework can learn from the homogeneity of target nodes and context nodes in a discriminative manner during the following hypergraph pretraining.
On the one hand, the information from pretraining is preserved in the target node embeddings and directly transferred to the downstream task. 
On the other hand, although context nodes are not in the downstream task, they provide the necessary information for the model to capture broader contextual patterns that are not limited to the nodes seen during the downstream task. The relation between target nodes can depend on the similarity among context nodes in pretraining, which is reflected by the context hypergraph. For example, two target nodes related to two context nodes are similar if the two context nodes are similar.

Similarly, a target hypergraph $\mathcal{H}_\text{t}' = (\mathcal{V}_\text{t},\mathcal{E}_\text{t}')$ and a context hypergraph $\mathcal{H}_\text{c}' = (\mathcal{V}_\text{c}',\mathcal{E}_\text{c}')$ are defined for the downstream task with their corresponding incidence matrices $\mathbf{H}_\text{t}'$ and $\mathbf{H}_\text{c}'$.
For both pretraining and downstream tasks, the isolation between the two hypergraphs separates the information propagation paths. In this way, the information of context nodes will not be directly aggregated to the target nodes through hyperedges in hypergraph learning, and vice versa. This design prevents the over-smoothing issue~\cite{oversmooth1, oversmooth2} among target and context nodes, preventing the representations of target nodes from converging to the same values, especially within dense graph data.

% In addition, besides the basic types, hyperedges can be further customized if additional information is available in specific scenarios. For instance, all items in the same category or all papers under the same topic can be connected by one hyperedge. Furthermore, a hyperedge can be used to connect arbitrary nodes as the objectives of an instruction.
% In this paper, we build hyperedges based on the graph structure and whether the embeddings of its connected nodes are transferred to the downstream task, leaving more various constructions in future work.

% \begin{figure}[h]
% \centering
%     \includegraphics[width=\linewidth]{figures/hyperedge.pdf}
%   \caption{Construction of hyperedges in target and context hypergraphs}
%   \label{fig:construction}
% \end{figure}
\subsection{Prompt Hypergraph Convolution}
In hypergraph learning, hypergraph convolution is prevalently used for information propagation across nodes in hypergraphs. Nonetheless, current hypergraph convolution methods focus on the direct aggregation of structural information from neighbors connected by hyperedge. To integrate the information from instructions into the hypergraph convolution process, we devise a novel Prompt Hypergraph Convolution layer. First, we construct the hyperedge information based on connected nodes. This step is denoted as \textit{hyperedge initialization} of the PHC layer. Then, a \textit{information prompting} step is proposed to endow the fusion of hyperedge information from resources other than the hypergraph structure, such as instructions. Finally, fused information is aggregated back to update the node embeddings in the \textit{node update} step. To be more concrete, we present the calculation details of a PHC layer in the target hypergraph. The illustration is in Figure~\ref{fig:IHP}(c).

\subsubsection{Hyperedge Initialization}
In the target hypergraph $\mathcal{H}_\text{t}$, target nodes are connected by target-target and target-context hyperedges. Therefore, we initialize the representation of a hyperedge by aggregating embeddings of all nodes connected by this hyperedge. The initialized representation of a hyperedge $\epsilon\in \mathcal{E}_\text{t}$ is formulated as follows:
\begin{equation}\label{eq:initial}
    \mathbf{h}_\epsilon^{(l)} = \frac{1}{|\mathcal{N}_\epsilon|}\sum_{v\in\mathcal{N}_\epsilon}\mathbf{e}_{t}^{(l)},
\end{equation}
where $\mathbf{h}_\epsilon^{(l)}$ represents initialized embedding of the hyperedge $\epsilon$, $\mathcal{N}_\epsilon$ is the set of target nodes connected by this hyperedge, and $\mathbf{e}_{t}^{(l)}$ is the embedding of those target nodes as input of the $l$-th PHC layer.
We use mean-pooling as the information aggregation method to aggregate structural information from nodes to hyperedges.
\subsubsection{Information Prompting}
To enable the framework to capture high-order relations under the guidance of instructions during hypergraph convolution, we fuse the initialized representation of a hyperedge with prompted information. With the prompt embedding of a hyperedge $\epsilon$ denoted as $\mathbf{p}_\epsilon$, we fuse $\mathbf{h}_\epsilon^{(l)}$ with $\mathbf{p}_\epsilon$ by addition:
\begin{equation}\label{eq:info prompt}
    \mathbf{q}_\epsilon^{(l)} = \mathbf{h}_\epsilon^{(l)} + \gamma\mathbf{p}_{\epsilon}^{(l)},
\end{equation}
where $\mathbf{q}_\epsilon^{(l)}$ is the fused hyperedge embedding, and $\gamma$ is a scalar hyperparameter to control the prompting intensity. 

With instructions accessible, the prompt embedding can be obtained based on text information from instructions, as shown in Figure~\ref{fig:IHP}(b) and~\ref{fig:IHP}(c).
If the instruction information is not available, the flexibility of this PHC layer allows other information to be prompted to the hyperedge. For example, a context-target hyperedge in the context hypergraph is converted from a target node. Hence, we can adopt prompt embeddings from the target node embeddings for a context hypergraph without instructions, i.e., $\mathbf{P}^{(l)}=\mathbf{E}_\text{t}^{(l+1)}$, as shown in Figure~\ref{fig:IHP}(a). The target node embeddings $\mathbf{E}_\text{t}^{(l+1)}$ are output by the $l$-th PHC layer applied on the target hypergraph, after node update in Eq.~(\ref{eq:node update}).

\subsubsection{Node update} We update the target node embedding by aggregating the fused embeddings from all its connected hyperedges.
% the updated feature of the hyperedge is propagated back to its connected group node as Fig.~\ref{fig:framework}b(3) shows. 
For a target node $v_t$, the aggregation step is formulated as follows:
\begin{equation}\label{eq:node update}
    \mathbf{e}_t^{(l+1)} = \frac{1}{|\mathcal{N}_t|}\sum_{\epsilon\in\mathcal{N}_t}\mathbf{q}^{(l)}_{\epsilon},
\end{equation}
where $\mathbf{e}_{t}^{(l+1)}$ denotes the output target node embedding, $\mathcal{N}_t$ is the set of hyperedges connected to this target node, and $\mathbf{q}^{(l)}_{\epsilon}$ is the fused embedding of its connected hyperedges from Eq.~(\ref{eq:info prompt}).

\subsubsection{PHC in Matrix Form}
To offer a holistic view
of convolution, we formulate the matrix form of prompt hypergraph convolution (equivalent to Eq.~(\ref{eq:initial})-(\ref{eq:node update})) as:
\begin{equation}\label{thc}
\begin{aligned}
    \mathbf{E}_{\text{t}}^{(l+1)} &= \text{PHC}(\mathbf{E}_\text{t}^{(l)},\mathbf{H}_\text{t},\gamma\mathbf{P}^{(l)})\\
    &= \mathbf{D}^{-1}\mathbf{H}_\text{t}\cdot
   (\mathbf{B}^{-1}\mathbf{H}_\text{t}^\top\mathbf{E}_{\text{t}}^{(l)}+ \gamma\mathbf{P}^{(l)}),
\end{aligned}
\end{equation}
where  $\mathbf{E}_{\text{t}}^{(l)}$ is the input embedding from $l$-th layer, $\mathbf{H}_\text{t}$ is the incidence matrix, $\mathbf{P}^{(l)}$ denotes prompt embeddings for the hyperedges in the hypergraph, and  $\mathbf{E}_{\text{t}}^{(l+1)}$ is the output. $\mathbf{D}$ is the degree matrix of nodes and $\mathbf{B}$ is degree matrix of hyperedges for normalization. 
PHC layer is a more general version of hypergraph convolution, which degrades to existing hypergraph convolution~\cite{mhcn} with prompting intensity $\gamma = 0$.

\subsection{Instruction-based Prompt Embedding}
To provide explicit guidance for hypergraph learning, instruction-based prompt embeddings are constructed based on available task-related information and fed into the PHC layer for further information propagation. Each instruction is expected to assist in capturing the relationship between the nodes connected by one hyperedge. Hence, the instructions and the formulated hyperedges are designed in one-to-one correspondence. Instructions can be formulated from task descriptions, node descriptions, or a concatenation of both. 

We show an example of leveraging the concatenation of task and node descriptions as instructions in Figure~\ref{fig:IHP}. Given that target-context hyperedges are context nodes in the original graph, we concatenate the node descriptions $\mathcal{A}_c$ and the descriptions $\mathcal{T}_c$ of tasks related to these nodes as instructions $\mathcal{I}_c$, where $|\mathcal{I}_c|=|\mathcal{E}_{t,c}|$. 
In this paper, descriptions of context nodes are obtained from the dataset, and the task descriptions are manually created for different tasks. For instance, if target nodes are readers and context nodes are history books in pretraining, the task description can be formulated as 'promote history books to readers.' In other words, the task description consists of node information (e.g., readers) and/or other task information (e.g., promote or classify). 

After the construction of instructions, a pretrained sentence transformer~\cite{minilm} is deployed to encode the instructions as text embeddings $\mathbf{C}\in \mathbb{R}^{|\mathcal{I}_c|\times d'}$, which are frozen in both pretraining and downstream tasks. A linear layer is applied to adaptively learn the transformation from text embeddings to prompt embeddings:
\begin{equation}
    \mathbf{P}=\mathbf{C}\cdot\mathbf{W}_\text{p}+\mathbf{b}_\text{p}
\end{equation}
where $\mathbf{W}_\text{p}\in \mathbb{R}^{d'\times d}$ and $\mathbf{b}_\text{p}\in \mathbb{R}^d$ are the learnable weight matrix and bias for prompt transformation. And $\mathbf{P}$ is the output prompt embedding, which prompts the target-context hyperedges with instruction information in every PHC layer for the target hypergraph. 

\subsection{Pretraining and Optimization}
% \subsubsection{Pretraining task}
% The link prediction task is employed in our pretraining stage for optimization. Link prediction is widely recognized as an effective pretraining task, given the abundance of links present in large-scale graph data~\cite{graphprompt,gppt,gptgnn}. 
% In addition, benefiting from the flexibility of hyperedge construction, node classification tasks and graph classification tasks can also be reformulated as link prediction tasks on hypergraphs as below:
% \begin{itemize}[leftmargin=*]
%     \item \textbf{Node Classification}. To predict categories of nodes, each category is used as a hyperedge to connect the nodes within such category. Next, representations of both nodes and hyperedges are obtained through hypergraph convolution or attention.
%     In the end, we can predict the connection between nodes and categories as hyperedges as a link prediction task. 
%     \item \textbf{Graph Classification}. Similar to node classification, all nodes in a graph are connected to hyperedges corresponding to graph categories. After hypergraph learning, the link between nodes and hyperedges can be predicted based on their representations.
% \end{itemize}
\subsubsection{Prediction and optimization}
The link prediction task is employed in our pretraining stage for optimization. Link prediction is widely recognized as an effective pretraining task, given the abundance of links present in large-scale graph data~\cite{graphprompt,gppt,gptgnn}. We use the inner product of two nodes as the link prediction score. For example, in Figure~\ref{fig:IHP}(a), the link prediction score $y_{t,c}$ between a target node and a context node in pretraining is formulated as:
\begin{equation}
    y_{t,c}=\mathbf{e}_{t}\cdot\mathbf{e}_{c}
\end{equation}
where $\mathbf{e}_{t}$ and $\mathbf{e}_{c}$ are the final embeddings of the target node and the context node after PHC layers. We remove the superscripts for simplicity. Then, we adopt the pairwise BPR loss~\cite{bprloss} to optimize the prediction:
\begin{equation}
    \mathcal{L}_{bpr}=\sum\limits_{(u,v,v^-)\in \mathcal{D}} -\log\sigma(\hat{y}_{u,v}-\hat{y}_{u,v^-}) + \lambda_\Theta\|\Theta\|^2_2,
\end{equation}
where $\mathcal{D}=\{(u,v,v^-)|v\in G^{+}_{u}, v^-\in G\backslash G^{+}_{u}\}$ is the training data, and positive set $G^{+}_{u}$ contains all nodes connected to node $u$. All parameters $\Theta$ in the pretraining is regularized by $\lambda_\Theta$. Adam~\cite{adam14} is used as the optimizer.

\subsubsection{Instruction-based finetuning}
The output $\Theta$ of the pretraining stage is the optimal target node embeddings $\mathbf{E}_\text{t}$ and all parameters $\Theta_\text{p}=\{\mathbf{W}_\text{p},\mathbf{b}_\text{p}\}$ in the prompt layer. This pretrained parameter $\Theta$ is used to initialize the target nodes and prompt layer in the downstream task.  In the downstream task, we freeze $\Theta_\text{p}$ in the prompt transformation and further finetune the embeddings of pretrained target nodes and unseen context nodes with downstream instructions. The reasons are twofold. Firstly, freezing this transformation ensures that the model responds to instructions
consistently in both pretext and downstream tasks. Since the task-related information is inherently encapsulated in instructions, it is unnecessary to finetune the transformation for learning the task-related information. Second, updating the unseen context nodes with the pretrained target nodes ensures that their features and relationships are captured in the context of the overall graph. This allows for dynamic adaptation of node representations to the specific context of the downstream task, guaranteeing that all nodes are represented according to the downstream instructions.

To address the catastrophic forgetting issue in finetuning, we deploy an adaptation intensity coefficient $\lambda_\text{t}$ for the learning rate $\eta_\text{t}$ of target node embeddings. We denote the learning rate of context node embeddings denoted as $\eta_\text{c}$ and set $\eta_\text{t} = \lambda_\text{t}\cdot\eta_\text{c}$. A lower learning rate for target nodes prevents the framework from forgetting the prior knowledge preserved in the target node embeddings. In this way, the pretrained model achieves a balance between retraining prior knowledge and adapting effectively for downstream tasks.

\subsubsection{Efficiency}
Our framework is efficient in both space and time complexity. Node embeddings $\mathbf{E}$, prompt transformation matrix $\mathbf{W}_\text{p}$ and prompt transformation bias $\mathbf{b}_\text{p}$ are the only learnable parameters in IHP. Given the size of pretrained text embeddings as $\mathcal{O}(|V|)$ for pretraining and $\mathcal{O}(|V'|)$ for finetuning, the total space complexity is $\mathcal{O}(|V|d+|V|d'+d'd)$ for pretraining and $\mathcal{O}(|V'|d+|V'|d'+d'd)$ for finetuning, where $\forall x\in\{d,d'\},\forall y\in\{|V|,|V'|\}, x\ll y$. The time complexity depends on the interaction between hyperedges and nodes, and the number of PHC layers $L$. For the four types of hyperedges constructed in IHP, the number of interactions is $\mathcal{O}(|E|)$ for pretraining and $\mathcal{O}(|E'|)$ for finetuning. Therefore, the time complexity is only $\mathcal{O}(L|E|d+d'd)$ for pretraining and $\mathcal{O}(L|E'|d)$ for finetuning, where $\forall x\in\{d,d'\},\forall y\in\{|E|,|E'|\}, L\ll x\ll y$. In contrast, a typical graph model GCN~\cite{GCN16} needs $\mathcal{O}(L|V|d^2+L|E|d+|V|d)$ time with $|V|$ nodes and $|E|$ edges.

\begin{table}\caption{The statistics of datasets.}\label{tab:dataset}
\begin{tabular}{l|l|l|l}
\hline
\hline
Dataset             & Goodreads-P & Goodreads-H & Amazon  \\ \hline
\# nodes            & 69,511      & 220,704     & 362,900 \\
\# edges            & 370,326     & 1,673,926   & 726,531 \\
\# target nodes     & 10,000      & 10,000      & 22,899  \\
\# pretrained nodes & 52,698      & 163,752     & 342,738 \\
\# pretrained edges & 271,344     & 1,407,108   & 665,695 \\
\# node descriptions     & 59,511      & 210,704     & 340,001 \\ \hline
\end{tabular}
\end{table}

\begin{table*}[]\caption{Link prediction performance. The best and second-best results are
in boldface and underlined, respectively.}\label{tab:overall performance}
\begin{tabular}{l|cccc|cccc|cccc}
\hline
Dataset      & \multicolumn{4}{c|}{Goodreads-P}    & \multicolumn{4}{c|}{Goodreads-H}                    & \multicolumn{4}{c}{Amazon}                    \\ \hline
Metric       & R@10         & R@20   & N@10    & N@20   & R@10         & R@20         & N@10         & N@20         & R@10   & R@20   & N@10         & N@20         \\ \hline
LightGCN     & 0.2427       & 0.2958 & 0.1684  & 0.1846 & 0.0956       & 0.1111       & {\ul 0.0818} & {\ul 0.0869} & 0.0661 & 0.0828 & 0.0497       & 0.0545       \\
SGL          & 0.2365       & 0.2935 & 0.1637  & 0.1808 & {\ul 0.0960} & 0.1108       & 0.0810       & 0.0856       & 0.0683 & 0.0812 & 0.0494       & 0.0530       \\
DirectAU     & 0.2106       & 0.3152 & 0.1721  & 0.2011 & 0.0933       & 0.1084       & 0.0807       & 0.0861       & 0.0689 & 0.0808 & 0.0599       & 0.0630       \\
HCCF         & 0.2427       & 0.3175 & 0.1884  & 0.2092 & 0.0721       & 0.1062       & 0.0603       & 0.0735       & 0.0682 & 0.0809 & 0.0605       & 0.0639       \\
DHCF         & 0.2440       & 0.2995 & 0.1702  & 0.1846 & 0.0735       & 0.0919       & 0.0569       & 0.0636       & 0.0299 & 0.0399 & 0.0177       & 0.0207       \\
GraphFormers & 0.2147       & 0.2770 & 0.1388  & 0.1583 & 0.0889       & {\ul 0.1187} & 0.0757       & 0.0854       & 0.0401 & 0.0496 & 0.0285       & 0.0313       \\ \hline
AttriMask    & -            & -      & -       & -      & 0.0640       & 0.0960       & 0.0489       & 0.0600       & 0.0741 & 0.0871 & {\ul 0.0633} & {\ul 0.0670} \\
GCC          & {\ul 0.2585} & 0.3185 & 0.1933  & 0.2117 & 0.0623       & 0.0947       & 0.0465       & 0.0584       & 0.0749 & 0.0905 & 0.0611       & 0.0656       \\
GraphMAE &
  0.2574 &
  {\ul 0.3220} &
  {\ul 0.1984} &
  {\ul 0.2178} &
  0.0678 &
  0.1020 &
  0.0557 &
  0.0678 &
  {\ul 0.0768} &
  {\ul 0.0920} &
  0.0585 &
  0.0627 \\ \hline
IHP &
  \textbf{0.2782} &
  \textbf{0.3380} &
  \textbf{0.2189} &
  \textbf{0.2351} &
  \textbf{0.1032} &
  \textbf{0.1319} &
  \textbf{0.0915} &
  \textbf{0.1005} &
  \textbf{0.0814} &
  \textbf{0.0980} &
  \textbf{0.0692} &
  \textbf{0.0738} \\ \hline
Improv.      & 7.63\%       & 4.99\% & 10.31\% & 7.93\% & 7.49\%       & 11.07\%      & 11.77\%      & 15.71\%      & 5.97\% & 6.59\% & 9.29\%       & 10.17\%      \\ \hline
\end{tabular}
\end{table*}
\section{Experiment}

\subsection{Experimental Setup}
\subsubsection{Datasets}

We conduct experiments on three real-world datasets: Goodreads-P,  Goodreads-H, and Amazon. Goodreads is a publicly available large-scale dataset including information about online readers and books on their shelves~\cite{dataset_goodreads}. We regard online readers as the target nodes for the two Goodreads datasets. For Goodreads-P, poetry books are used as context nodes for pretraining, and comics are used as downstream context nodes.  For Goodreads-H, history books and biographies are downstream context nodes, while pretraining context nodes contain books in the categories of children, young adults, mystery, thriller, and crime. For these two Goodreads datasets, we predict the link between readers and books. Amazon includes users and items purchased by them~\cite{dataset_amazon}. Users are regarded as target nodes, and instruments are used as context nodes in the downstream task. For pretraining in Amazon, context nodes are items in the categories of arts, crafts, sewing, grocery, gourmet food, office products, electronics, sports, outdoors, toys, and games. For Amazon, we predict the link between users and items. The main statistics of the three datasets are summarized in Table~\ref{tab:dataset}. For downstream link prediction tasks in all the datasets, we split $70\%$ of edges for training and the remaining $30\%$ for testing.

\subsubsection{Baselines}
% We compared IHP with the following link prediction methods as baselines:
% \begin{itemize}[leftmargin=*]
%     \item \textbf{LightGCN}~\cite{lightgcn20}. This is the state-of-the-art recommendation method based on GCN~\cite{GCN16} by removing feature transformation and nonlinear activation.
%     \item \textbf{SGL}~\cite{sgl}. This model performs contrastive learning on LightGCN to augment node representations by leveraging additional self-supervised signals.
%     \item \textbf{DirectAU}~\cite{auloss}. This graph learning method proposes a novel loss based on alignment and uniformity of embedding distribution for model optimization.
%     \item \textbf{HCCF}~\cite{hccf}. It employs hypergraph learning with contrastive learning to jointly capture local and global high-order relations.
%     \item \textbf{DHCF}~\cite{dhcf}. It employs residual connections to model hybrid multi-order correlations through jump hypergraph convolution.
%     \item \textbf{Graphformers}~\cite{graphformer}. This GNN-nested Transformer model encodes text and structure information iteratively. We use node descriptions as the input text information.
% \end{itemize}
We regard users as target nodes and items as context nodes in all the datasets. Thus recommendation methods, including LightGCN~\cite{lightgcn20}, SGL~\cite{sgl}, HCCF~\cite{hccf} and DHCF~\cite{dhcf}, are applied as strong baselines to prediction interactions between target and context nodes, besides baseline models designed for general graph tasks, such as DirectAU~\cite{auloss} and GraphFormers~\cite{graphformer}. As a pretraining framework, we also compare IHP with the following pretraining baselines:
\begin{itemize}[leftmargin=*]
    \item \textbf{AttriMask}~\cite{attrimask}. This method applies GNN to obtained node embeddings and then adds a linear model to predict masked attributes during pretraining. We use node categories as the masked attributes, so it is not suitable for Goodreads-P where all pretraining nodes are comics.
    \item \textbf{GCC}~\cite{gcc}. This self-supervised pretraining framework leverages contrastive learning with random walks to capture structure information in graphs.
    \item \textbf{GraphMAE}~\cite{graphmae}. This method applies a masked graph autoencoder for generative self-supervised graph pretraining.
\end{itemize}
For the pretraining baselines, we use the same finetuning stage as IHP, which is shown in Figure~\ref{fig:IHP}(a), with no instructions and instruction prompting intensity $\gamma=0$.
\subsubsection{Evaluation metric}
We evaluate the pretraining framework by ranking the test context nodes with all non-interacted target nodes during finetuning. Recall$@ \{10,20\}$ and NDCG$@\{10,20\}$ are adopted as evaluation metrics.  
\subsubsection{Implementations}
For all the datasets, the pretraining learning rate is set to $0.1$. The regularization parameter $\lambda_\Theta$ is set to 1e-7. 
The node embedding size is set to 64.
The learning rate to finetune context nodes is set to 0.1 for the two Goodreads datasets, and 0.01 for Amazon.
For the Goodreads-P/Goodreads-H/Amazon dataset, we set the number of PHC layers as 4/1/3, the prompting intensity in the target hypergraph as 1e-3/1e-4/1e-1, the prompting intensity in the context hypergraph as 1/1/1e-4, and the adaptation intensity as 1e-2/1e-3/10. We implement early stopping based on loss for pretraining and Recall@10 for finetuning. All the experiments are conducted on an 8GB RTX 2080. 
As Table~\ref{tab:time} shows, the framework can be pretrained and finetuned within one minute for all the datasets, benefiting from the compact size of learnable parameters and the high efficiency.
%Benefiting from the compact size of learnable parameters and the high efficiency, the total pretraining time of IHP is only 7.80 seconds for Goodreads-P, 5.60 seconds for Goodreads-H, and 23.79 seconds for Amazon. The finetuning time is only 10.61 seconds for Goodreads-P, 28.78 seconds for Goodreads-H, and 24.11 seconds for Amazon. 
Our implementation is available at \url{https://anonymous.4open.science/r/IHP-ECC8}.
\begin{table}\caption{The total training time of IHP.}\label{tab:time}
\begin{tabular}{l|l|l|l}
\hline
Dataset             & Goodreads-P & Goodreads-H & Amazon  \\ \hline
Pretraining time (s) & 7.80     & 5.60   & 23.79 \\
Finetuning time (s)  & 10.61    & 28.78  & 24.11  \\ \hline
\end{tabular}
\end{table}

\subsection{Performance on Link Prediction }
We show the overall comparison of link prediction performance in Table~\ref{tab:overall performance}. The best results are in boldface. We observe that the proposed IHP framework achieves the best results and outperforms all the baselines in the three datasets. 
We hypothesize these large and stable gains result from prior knowledge learned from pretraining under the explicit guidance of instructions. In addition, baseline pretraining frameworks achieve better performance than models without pretraining in Goodreads-P and Amazon. However, they are surpassed by other baselines in Goodreads-H. Since the pretrained edges in Goodreads-H are densest, and pretrained node types in it are fewer than in Amazon, the risk of overfitting for baseline pretraining frameworks is high in such a dataset. In contrast, instructions provide explicit information about certain tasks, which explains IHP's better generalization and adaptation for the downstream task.

\begin{table}[]\caption{Node classification performance.}\label{tab:classification}
\begin{tabular}{l|cc|cc}
\hline
Dataset      & \multicolumn{2}{c|}{Goodreads-H}    & \multicolumn{2}{c}{Amazon}         \\ \hline
Metric       & Acc.             & F1               & Acc.             & F1               \\ \hline
DirectAU     & 0.64494          & 0.64135          & 0.77066          & 0.77006          \\
HGNN         & 0.64924          & 0.64566          & {\ul 0.77315}    & {\ul 0.77302}    \\
GraphFormers & 0.64412          & 0.64042          & 0.76871          & 0.77148          \\ \hline
AttriMask    & {\ul 0.64951}    & {\ul 0.64626}    & 0.77298          & 0.77284          \\
GCC          & 0.64949          & 0.64620          & 0.77284          & 0.77271          \\
GraphMAE     & 0.64935          & 0.64610          & 0.77315          & 0.77301          \\ \hline
IHP          & \textbf{0.64962} & \textbf{0.64641} & \textbf{0.77341} & \textbf{0.77328} \\ \hline
\end{tabular}
\end{table}

\subsection{Node Classification and Performance}
We also evaluate IHP on node classification as the downstream task, optimized by cross-entropy loss~\cite{crossentropy}. Besides the previous baselines designed for general graph learning, HGNN~\cite{hgnn} leveraging hypergraph convolution for classification is also used as a baseline. For Goodreads-H, books in the categories of children and young adults are pretraining context nodes, and history, biography, mystery,
thriller and crime books are used in downstream classification. For Amazon, items in the categories of arts, crafts, sewing, grocery, gourmet food and office products are used for pretraining context nodes, and items in the categories of electronics, sports, outdoors, toys, games and instruments are used in downstream classification. We split $70\%$ of nodes for training and the remaining $30\%$ for testing. Accuracy and F1 scores are adopted as evaluation metrics. To prevent data leakage, we removed node category information from task descriptions in the instructions. The parameters remain the same as the settings for the link prediction.

The comparison is shown in Table~\ref{tab:classification}. The performance verifies the efficacy of IHP in node classification tasks. The pretraining framework baselines have better performance than other baselines in most cases, which indicates that the node features can be accurately captured with the prior knowledge learned during pretraining. HGNN surpasses all other non-pretrained baselines, showing the benefit of modeling high-order relations by hypergraph learning. In addition, with inconsistency in pretraining and finetuning structures, pretrained baselines slightly degrade the downstream performance of hypergraph learning in Amazon, compared with HGNN. However, IHP with frozen prompt transformation maintains the consistency between pretraining and finetuning stages, and leverages instructions to offer information on downstream node features, which is a more comprehensive pretraining framework for node classification.

\subsection{Ablation Study}
\begin{table}[]\caption{Performance with different text encoders.}\label{tab:PLM}
\resizebox{0.48\textwidth}{!}{%
\begin{tabular}{l|cccccc}
\hline
Dataset & \multicolumn{2}{c}{Goodreads-P}      & \multicolumn{2}{c}{Goodreads-H}     & \multicolumn{2}{c}{Amazon}         \\ \hline
Metric    & R@10   & N@10   & R@10   & N@10   & R@10   & N@10   \\ \hline
w/o PLM  & 0.2717 & 0.1974 & 0.0944 & 0.0824 &0.0545 & 0.0406 \\
Instructor~\cite{instructor}  & 0.2763 & 0.2225 & 0.1019 & 0.0905 &0.0787 & 0.0662 \\
%Instructor-xl~\cite{instructor} & 0.2765 & 0.2226 &0.1020& 0.0900 & 0.0794 & 0.0671\\
GTR~\cite{gtr} & 0.2764 & {\ul 0.2226} & 0.1020 & 0.0900 & 0.0793 & 0.0665 \\
S-BERT~\cite{sentenceBERT}& {\ul 0.2765} & 0.2218 & {\ul 0.1024} & 0.0903 & \textbf{0.0831} & \textbf{0.0698} \\ 
E5~\cite{e5}& 0.2764 & \textbf{0.2227} & 0.1023 & {\ul 0.0909} & 0.0808 & 0.0675\\ 
MiniLM~\cite{minilm}  & \textbf{0.2782} & 0.2189 & \textbf{0.1032} & \textbf{0.0915} &{\ul 0.0814} & {\ul 0.0692}
\\ \hline
\end{tabular}%
}
\end{table}
\subsubsection{PLMs as text encoders.} To quantify the contribution of PLMs as the text encoder in our framework, we compare the link prediction performance of IHP with different PLMs and without PLM in Table~\ref{tab:PLM}. For the variant of IHP without PLM, we randomly initialize the text embedding as the input of the prompt layer. It is apparent that employing PLM to encode instructions leads to significant enhancement in all the datasets. The results also imply that the benefits of different PLMs depend on the data distribution of the certain dataset. In IHP, we use MiniLM as the sentence encoder because of its stable performance in all the datasets.

\subsubsection{Task information in Instructions.} 
\begin{figure}[]
    \begin{subfigure}{0.156\textwidth}
    \includegraphics[width=\textwidth]{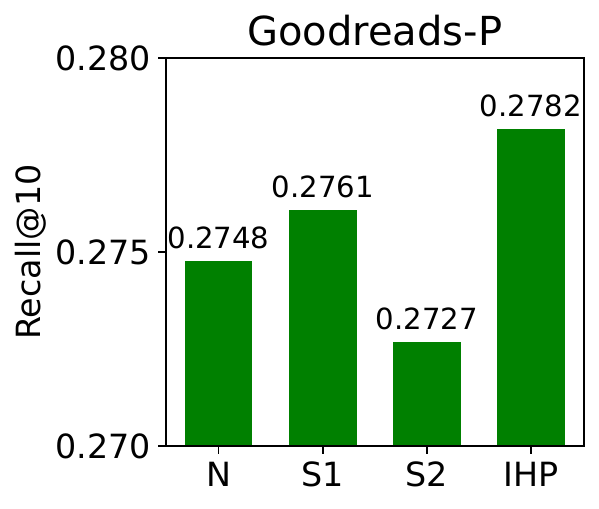}
    \end{subfigure}
    % \hspace{-2mm}
    \hfill
    \begin{subfigure}{0.156\textwidth}
    \includegraphics[width=\textwidth]{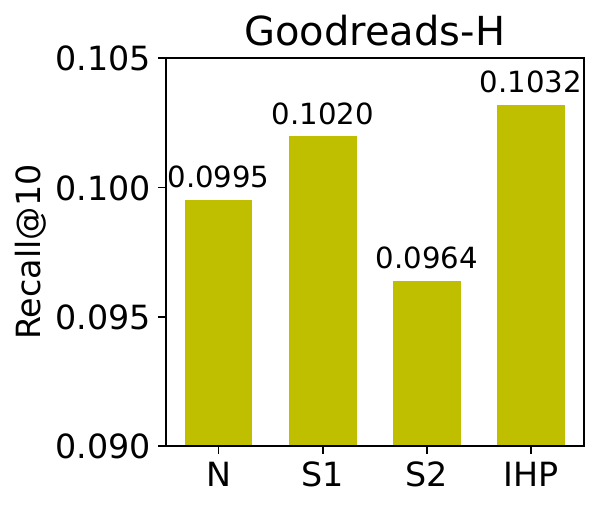}
    \end{subfigure}
    % \hspace{-2mm}
    \hfill
    \begin{subfigure}{0.156\textwidth}
    \includegraphics[width=\textwidth]{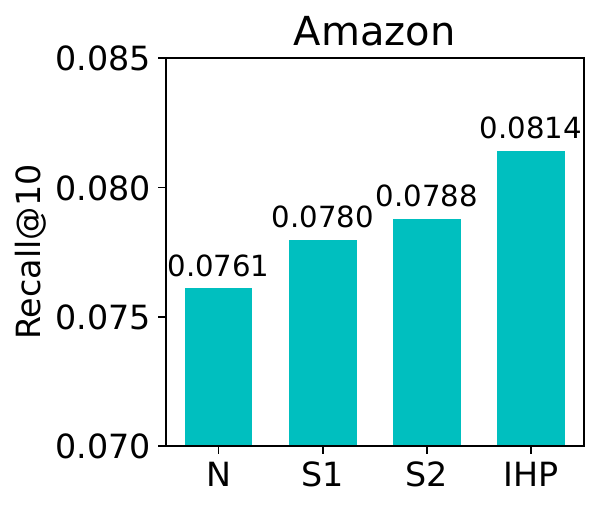}
    \end{subfigure}
        \caption{Performance of IHP w.r.t. different instruction constructions.}
  \label{fig:task_info}
\end{figure}

Besides node descriptions, the instructions in IHP also contain task descriptions (e.g., "promote electronics to the reader"). In Figure~\ref{fig:task_info}, We demonstrate the performance of IHP with different ways of leveraging task information as a part of the instruction. In addition to concatenating node descriptions and task information for text encoding, we investigate three other variants of IHP: (N) Without text-based task information, only node descriptions are encoded as text embeddings and then concatenated to a learnable task vector for each task before prompt transformation; 
(S1) node descriptions and task information are encoded by the PLM separately, and then concatenated for prompt transformation;
(S2) same as S1, but the encoded task embeddings skip both prompt transformation and hypergraph convolution. Instead, they are concatenated to corresponding node embeddings after PHC layers for the final prediction.

We observe that IHP performs the best on three datasets. This justifies the superiority of GTGS, which simultaneously encodes node and task information inside instructions. The poor performance of N verifies the limitation of learnable vectors compared to text-based task descriptions in instruction construction. S1 allows task information to participate in information propagation through PHC layers, which results in better performance than S2 in Goodreads datasets. When tasks become diverse in Amazon, S2 allocating task information to corresponding nodes can more accurately capture dependencies among nodes. However, separately encoding node descriptions and task information ignores the semantic connection between them, so both S1 and S2 are worse than IHP. 

\subsubsection{Instruction-based finetuning settings.} The prior knowledge is preserved in target node embeddings $\mathbf{E}_\text{t}$ and parameters of prompt layer $\Theta_\text{p}$ in the pretraining stage, and transferred to the downstream tasks. Hence, we conducted an ablation study to investigate the influence of different finetuning settings on downstream performance. For instruction-based finetuning in IHP, we freeze the prompt layer and update the target node embeddings with context node embeddings during downstream tasks. We compare IHP with the other four variants of finetuning settings: (i) IHP-TR with $\mathbf{E}_\text{t}$ randomly initialized in finetuning; (ii) IHP-TF with pretrained $\mathbf{E}_\text{t}$ frozen in finetuning; (iii) IHP-PR with $\Theta_\text{p}$ randomly initialized in finetuning; (iv) IHP-PU with pretrained $\Theta_\text{p}$ updated by the objective in finetuning.

\begin{figure}[]
    \begin{subfigure}{0.156\textwidth}
    \includegraphics[width=\textwidth]{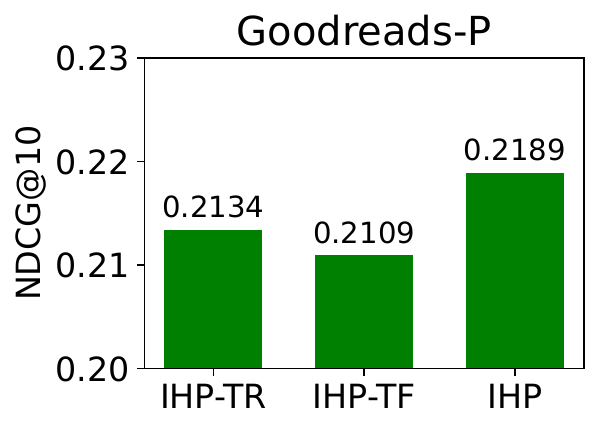}
    \end{subfigure}
    % \hspace{-2mm}
    \hfill
    \begin{subfigure}{0.156\textwidth}
    \includegraphics[width=\textwidth]{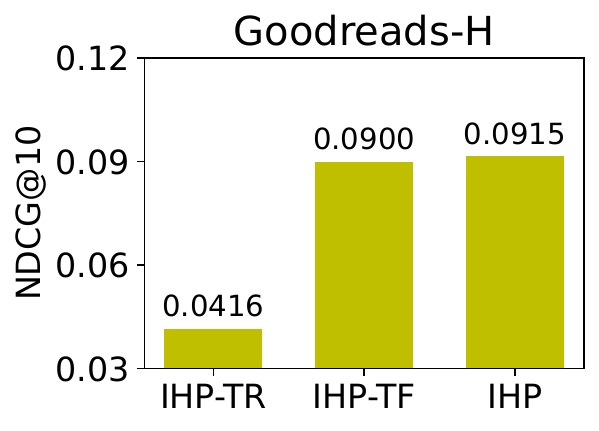}
    \end{subfigure}
    % \hspace{-2mm}
    \hfill
    \begin{subfigure}{0.156\textwidth}
    \includegraphics[width=\textwidth]{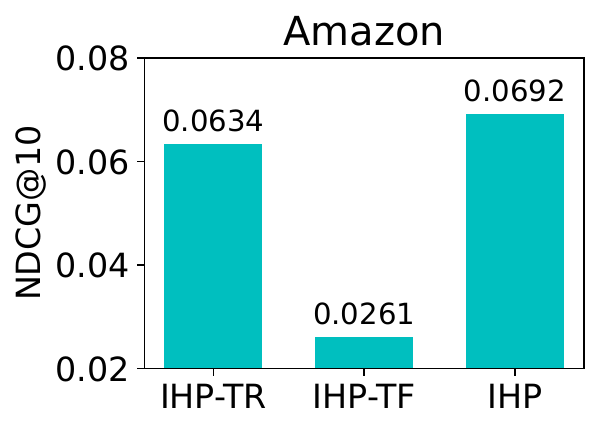}
    \end{subfigure}
        \caption{Performance of IHP w.r.t. random, frozen or updated target node embeddings during the finetuning stage.}
  \label{fig:target}
\end{figure}

\begin{figure}[]
    \begin{subfigure}{0.156\textwidth}
    \includegraphics[width=\textwidth]{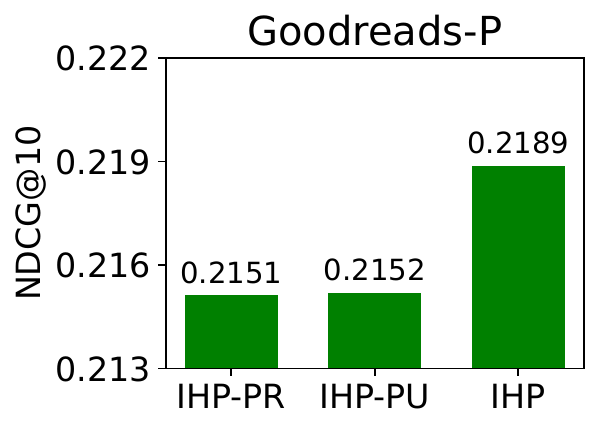}
    \end{subfigure}
    % \hspace{-2mm}
    \hfill
    \begin{subfigure}{0.156\textwidth}
    \includegraphics[width=\textwidth]{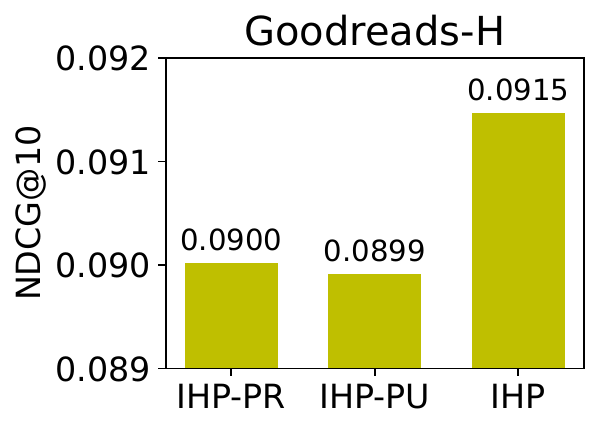}
    \end{subfigure}
    % \hspace{-2mm}
    \hfill
    \begin{subfigure}{0.156\textwidth}
    \includegraphics[width=\textwidth]{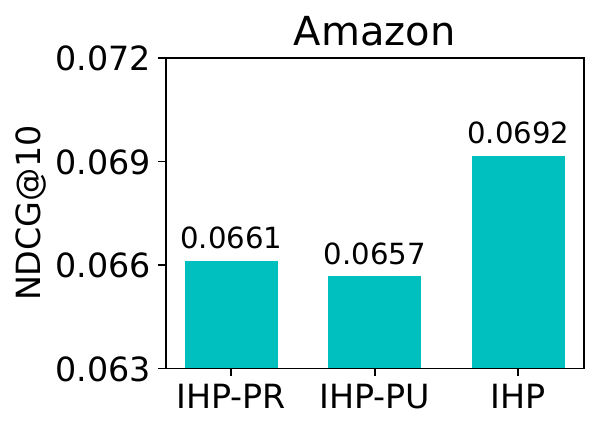}
    \end{subfigure}
        \caption{Performance of IHP w.r.t. random, updated or frozen prompt layer during the finetuning stage.}
  \label{fig:prompt}
\end{figure}

The comparison between the performance of IHP and its variants in Figures~\ref{fig:target} and~\ref{fig:prompt} confirms the advantage of our instruction-based finetuning paradigm. Freezing target node embeddings reduces the flexibility and limits the adaptability to downstream data for the pretraining framework. This issue becomes more pronounced with the growing discrepancy between pretrained and downstream data. As shown in Figure~\ref{fig:target}, IHP-TF with frozen target node embeddings can hardly adapt to the downstream instrument data from the pretrained item data from diverse categories in Amazon. Additionally, updating the prompt layer scarcely improves the downstream performance, compared with using randomly initialized parameters in Figure~\ref{fig:prompt}. On the contrary, IHP with a frozen prompt layer maintains the consistency of the model's response to instructions during pretraining and finetuning, and thus archives a better performance.

\begin{figure*}[h]
    \begin{subfigure}{0.33\textwidth}
    \includegraphics[width=\textwidth]{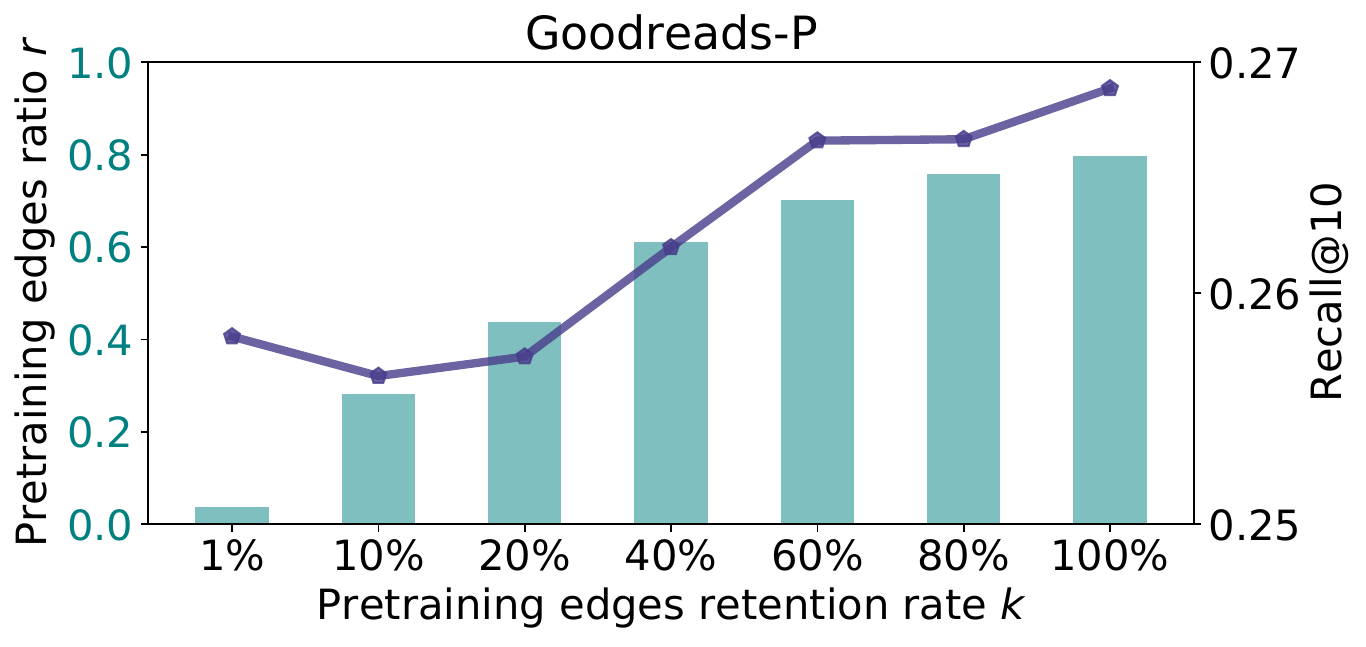}
    \end{subfigure}
    \hspace{-8mm}
    \hfill
    \begin{subfigure}{0.33\textwidth}
    \includegraphics[width=\textwidth]{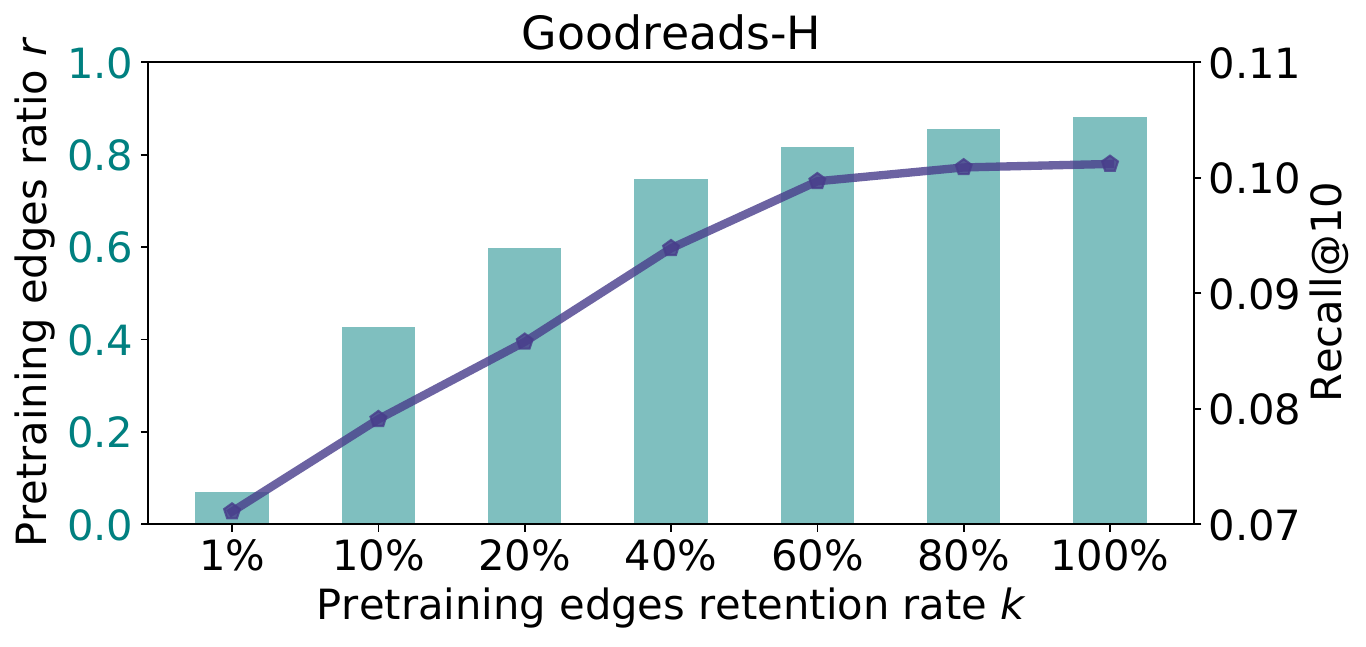}
    \end{subfigure}
    \hspace{-8mm}
    \hfill
    \begin{subfigure}{0.33\textwidth}
    \includegraphics[width=\textwidth]{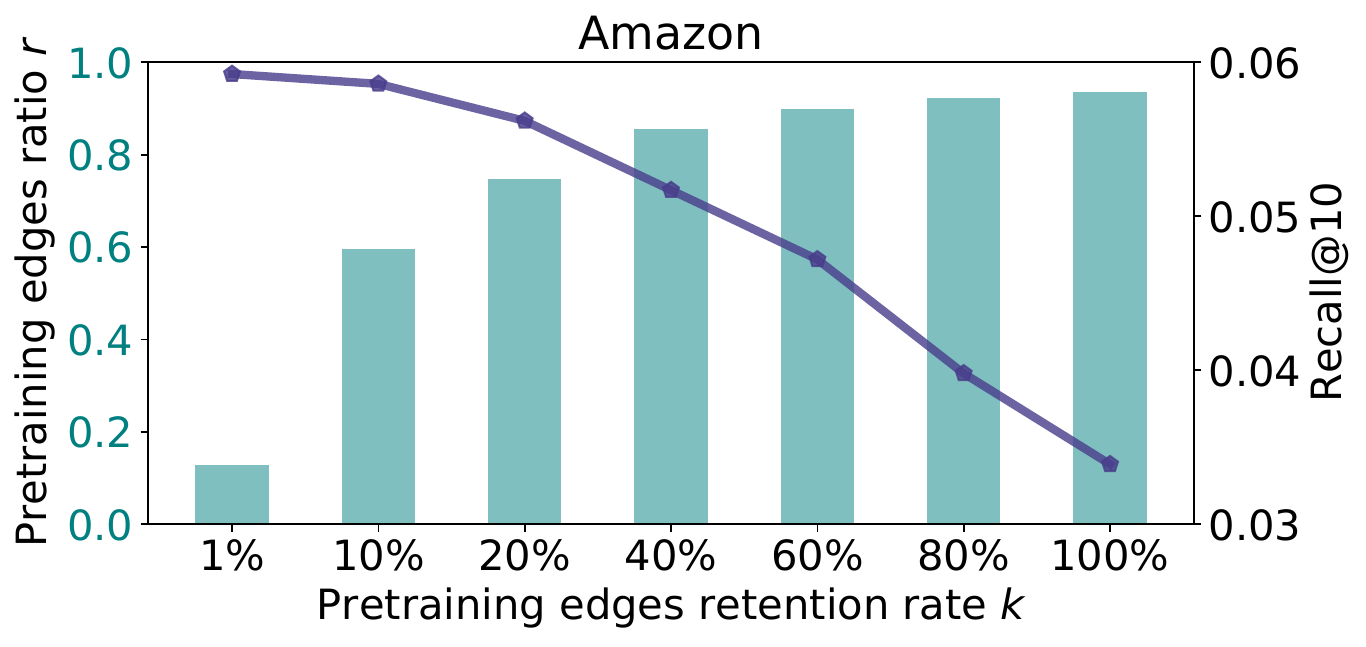}
    \end{subfigure}
        \caption{IHP finetuned with pretrained data added if the learning rates of target and context node embeddings are the same, i.e., $\lambda_\text{t}=1$. The curve represents the performance, and the bars denote the ratios of the edges added from pretraining in finetuning.}
  \label{fig:forgetting}
\end{figure*}

\begin{figure}[]
    \begin{subfigure}{0.156\textwidth}
    \includegraphics[width=\textwidth]{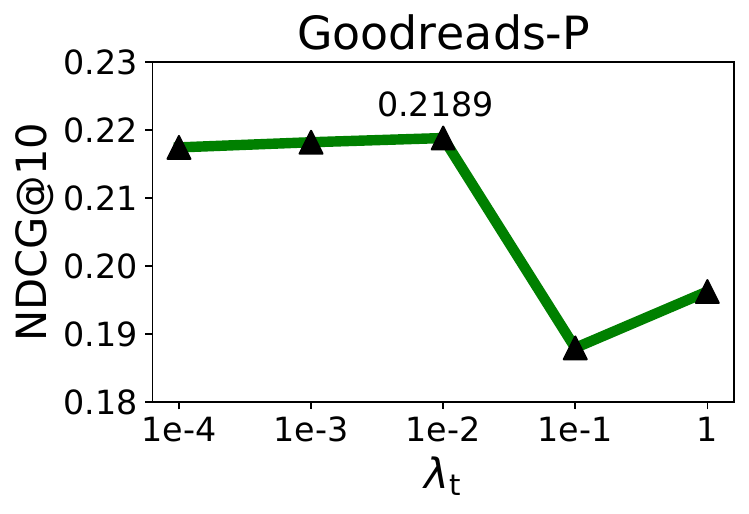}
    \end{subfigure}
    % \hspace{-2mm}
    \hfill
    \begin{subfigure}{0.156\textwidth}
    \includegraphics[width=\textwidth]{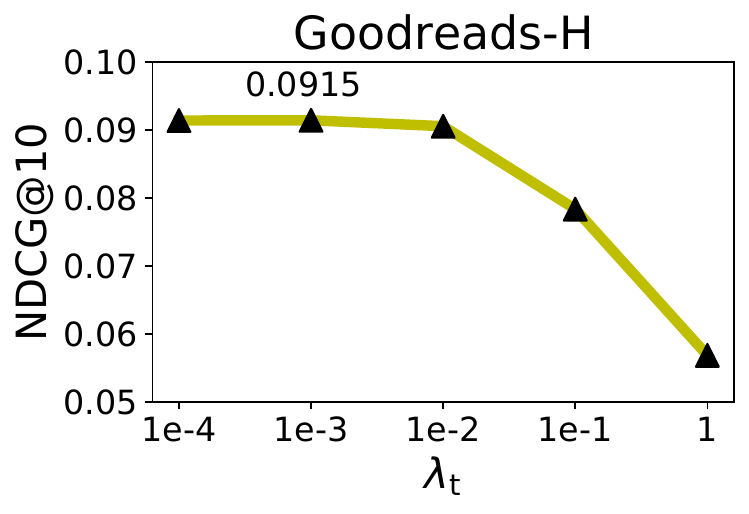}
    \end{subfigure}
    % \hspace{-2mm}
    \hfill
    \begin{subfigure}{0.156\textwidth}
    \includegraphics[width=\textwidth]{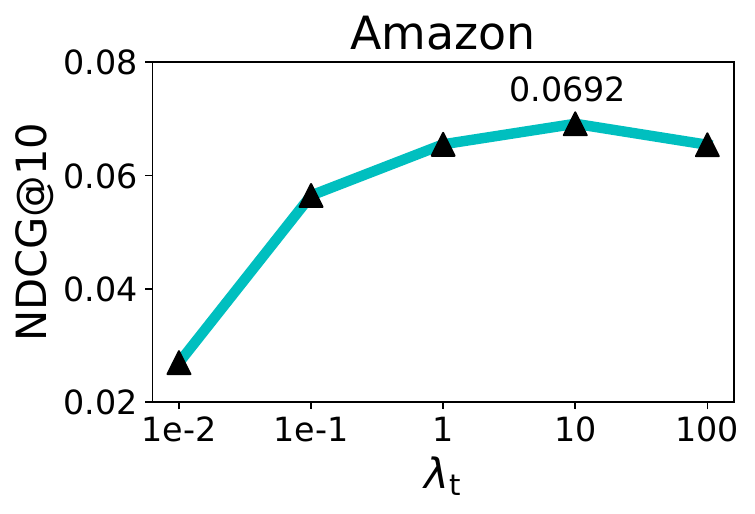}
    \end{subfigure}
        \caption{Performance of IHP w.r.t. $\lambda_\text{t}$.}
  \label{fig:decay}
\end{figure}

\subsection{Inductive Learning Analysis}
To validate the effectiveness of IHP in dynamically evolving graph structures, we add inductive target nodes which account for $25\%$, $40\%$, and $50\%$ of the original target nodes into Amazon, Goodreads-P, and Goodreads-H datasets. The embeddings of inductive nodes are never trained in either pretraining or finetuning stages. Instead, the inductive nodes are connected to other downstream training nodes by hyperedges defined in Section~\ref{sec:hyperedge} and fed into IHP only for inference. Then, we evaluate the link prediction performance on those inductive nodes. In addition to the pretraining baselines~\cite{attrimask,gcc,graphmae}, we also compare IHP with GraphSage~\cite{graphsage} which is a general inductive graph learning method.

The results are demonstrated in Table~\ref{tab:induct}. One can observe that IHP exhibits the best performance on inductive nodes in all the datasets, especially in larger Goodreads-H and Amazon datasets. With prior knowledge from abundant pretrained data in these two datasets, pretraining baselines always perform better than scratch-training GraphSage. Nonetheless, they are worse than IHP, which better captures dependencies between inductive and trained nodes by directly prompting text-based instruction to hyperedges.
\begin{table}[]\caption{Performance on inductive nodes.}\label{tab:induct}
\resizebox{0.48\textwidth}{!}{%
\begin{tabular}{l|cccccc}
\hline
Dataset & \multicolumn{2}{c}{Goodreads-P}      & \multicolumn{2}{c}{Goodreads-H}     & \multicolumn{2}{c}{Amazon}         \\ \hline
Metric     & R@10   & N@10   & R@10   & N@10   & R@10   & N@10   \\ \hline
GraphSage  & 0.2415 & {\ul0.1988} & 0.0305 & 0.0210 & 0.0330 & 0.0166 \\
AttriMask  & -      & -      & 0.0605 & 0.0489 &0.0887 & 0.0505 \\
GCC        & 0.2601 & 0.1960 & 0.0589 & 0.0483 &{\ul0.0929} & {\ul0.0520} \\
GraphMAE   & {\ul0.2620} & 0.1981 & {\ul0.0606} & {\ul0.0503} &0.0869 & 0.0494 \\ \hline
IHP        & \textbf{0.2622} & \textbf{0.2002} & \textbf{0.0998} & \textbf{0.0783} &\textbf{0.0939} & \textbf{0.0672}
\\ \hline
\end{tabular}%
}
\end{table}

\subsection{Adaptation Intensity Analysis}
To maintain the balance between retaining prior knowledge and adapting for various downstream tasks, we deploy the adaptation intensity coefficient $\lambda_\text{t}$ to differentiate the learning rates of target and context node embeddings during finetuning. We demonstrate the impact of this coefficient on three datasets in Figure~\ref{fig:decay}. It becomes obvious that the performance is always suboptimal when the learning rates of target and context node embeddings remain the same, i.e., $\lambda_\text{t}=1$. For the two Goodreads datasets, a small $\lambda_\text{t}$ prevents the framework from forgetting the prior
knowledge during finetuning. For Amazon, a large $\lambda_\text{t}$ is required to endow the model with the higher adaptivity to the downstream data. 

We believe the difference of $\lambda_\text{t}$ in Goodreads and Amazon results from the discrepancy between the pretraining and downstream data in each dataset. For verification, we evaluate the model performance with the same learning rates of target and context nodes after merging the pretraining and downstream data.
To be specific, we randomly select a subset $E_\text{s}$ of edges from the set $E$ of all the pretraining edges, add them to the downstream data, and then finetune the model with $\lambda_\text{t}=1$ on the downstream tasks. In other words, the edges become $E_\text{s}\cup E'$ for the downstream tasks. The proportion of the selected edges in all the pretrained edges is defined as pretraining edges retention rate $k=|E_\text{s}|/|E|$. The ratio of these selected edges to all the downstream edges is denoted as pretraining edges ratio $r=|E_\text{s}|/(|E_\text{s}|+|E'|)$. Pretraining context nodes connected by $E_\text{s}$ are also added to the downstream task. In this way, the model is finetuned directly with the original pretraining data.

The results are demonstrated in Figure~\ref{fig:forgetting}. In the two Goodreads datasets, We observe that the downstream performance is generally improved with the pretraining data added to the downstream task. In Amazon, on the contrary, the performance is degraded after adding the pretraining data when $\lambda_\text{t}=1$. The reason is that all the pretraining and downstream context nodes are books in Goodreads, so the discrepancy between the pretraining and downstream data in Goodreads is minor compared to the discrepancy in Amazon, where pretraining and downstream context nodes are items in more diverse categories. That explains why a small $\lambda_\text{t}$ is necessary to prevent the framework from forgetting the prior knowledge in Goodreads, while a large $\lambda_\text{t}$ can help the pretrain model faster adapt to the significantly different downstream data in Amazon.
\section{conclusion}
In this paper, we propose a novel graph pretraining framework IHP, which applies text-based instructions to overcome the discrepancy between pretraining and downstream tasks. In IHP, we construct hyperedges to capture high-order relations among nodes under the guidance of instructions, and a PHC layer is introduced to integrate instructions into context-aware information propagation in hypergraph learning. We conduct extensive experiments and detailed analyses on three real-world datasets to verify the effectiveness of IHP. In the future, further research is needed to evaluate the scalability and generalization of IHP on a broader range of datasets. With potential additional information in specific scenarios, hyperedges can be further customized (e.g., one hyperedge connecting all nodes with the same attribute) or used to connect arbitrary nodes as the objectives of an instruction. Given such flexibility of hypergraph construction, future studies could explore how to extend the IHP framework into multi-task pretraining, where different hypergraphs could be designed for diverse pretraining tasks. 

\section{acknowledgements}
This work is supported in part by NSF under grant III-2106758. 
%%
%% The next two lines define the bibliography style to be used, and
%% the bibliography file.
\bibliographystyle{ACM-Reference-Format}
\bibliography{sample-base}

\end{document}